\documentclass[%
 reprint,
superscriptaddress,
 amsmath,amssymb,
 aps,
]{revtex4-2}

\usepackage{graphicx}
\usepackage{dcolumn}
\usepackage{bm}

\newcommand{\summ}{\Sigma_{\nu}}

\usepackage{lineno}
\usepackage{ulem}
\usepackage{color}
\usepackage{comment}
\usepackage{float}

\begin{document}

\preprint{APS/123-QED}

\title{Target Neutrino Mass Precision for Determining the Neutrino Hierarchy}

\author{Constance Mahony}
\email{constance.mahony.16@ucl.ac.uk}
 \affiliation{Department of Physics \& Astronomy, University College London, Gower Street, London WC1E 6BT, UK}
\author{Boris Leistedt}
\affiliation{Center for Cosmology and Particle Physics, Department of Physics, New York University, 726 Broadway, New York, NY 10003, USA}
\affiliation{NASA Einstein fellow}
\author{Hiranya V. Peiris}
\affiliation{Department of Physics \& Astronomy, University College London, Gower Street, London WC1E 6BT, UK}
\affiliation{The Oskar Klein Centre for Cosmoparticle Physics, Department of Physics, Stockholm University, AlbaNova, Stockholm, SE-106 91, Sweden}
\author{Jonathan Braden}
\affiliation{Department of Physics \& Astronomy, University College London, Gower Street, London WC1E 6BT, UK}
\affiliation{Canadian Institute for Theoretical Astrophysics, University of Toronto, 60 St. George St, Toronto, ON, M5S 3H8, Canada}
\author{Benjamin Joachimi}
 \affiliation{Department of Physics \& Astronomy, University College London, Gower Street, London WC1E 6BT, UK}
\author{Andreas Korn}
 \affiliation{Department of Physics \& Astronomy, University College London, Gower Street, London WC1E 6BT, UK}
\author{Linda Cremonesi}
 \affiliation{Department of Physics \& Astronomy, University College London, Gower Street, London WC1E 6BT, UK}
\author{Ryan Nichol}
 \affiliation{Department of Physics \& Astronomy, University College London, Gower Street, London WC1E 6BT, UK}

\date{\today}

\begin{abstract}
Recent works combining neutrino oscillation and cosmological data to determine the neutrino hierarchy found a range of odds in favour of the normal hierarchy. These results arise from differing approaches to incorporating prior knowledge about neutrinos. We develop a hierarchy-agnostic prior and show that the hierarchy cannot be conclusively determined with current data. The determination of the hierarchy is limited by the neutrino mass scale $\summ$ measurement. We obtain a target precision of $\sigma(\summ) = 0.014$ eV, necessary for conclusively establishing the normal hierarchy with future data.
\end{abstract}

\maketitle


\section{Introduction}

Particle physics and cosmology provide complementary information about neutrinos. Neutrino oscillation experiments have determined that neutrinos have mass, contrary to the Standard Model, and that there are three mass eigenstates \cite{deSalas:2018bym}. They have also measured two mass-squared splittings between these mass states.  However, the overall scale of the neutrino masses is unknown, as is the ordering of the two squared splittings --- the neutrino hierarchy. Cosmological data place a constraint on the sum of neutrino masses, which provides an overall scale and will help to distinguish the hierarchy \cite{Splitting_constraints}. Determining the neutrino hierarchy is key to further understanding the properties of the neutrino sector and theories of neutrino mass generation \cite{Patterson:2015xja}.

The two possible orderings of the neutrino mass states are the Normal Hierarchy (NH) and the Inverted Hierarchy (IH). The IH has a greater total mass. The minimum total mass for the NH and IH can be calculated by fixing the mass of the lightest state at zero and using current squared splitting measurements to calculate the mass of the other two states. In the NH configuration the minimum total mass is 0.06 eV, and in the IH configuration the minimum total mass is 0.1 eV. Recent cosmological measurements have placed a 95\% CL upper bound on the sum of the neutrino masses of 0.12 eV \cite{Planck2018}, which is tantalizingly close to the IH minimum mass. 

Motivated by these results, many works have performed joint analyses of neutrino oscillation and cosmology data to see if there is already sufficient evidence for the NH \cite{Simpson:2017qvj, Hannestad:2016fog, Gerbino:2016ehw, Gariazzo:2018pei, Heavens:2018adv, Long:2017dru, Vagnozzi:2017ovm}. The results of these analyses vary dramatically, producing relative odds favoring the NH over the IH ranging from 3:2 to 470:1. The main difference between these analyses is how they incorporate the state of knowledge about the neutrino hierarchy before taking any data into account: their choice of prior \cite{Heavens:2018adv, Gariazzo:2018pei}. Choosing an appropriate prior is difficult because a physically-motivated prior on the neutrino properties (whether they are the individual masses, squared splittings, mixing angles, etc) does not exist \cite{Long:2017dru}. Further, there is a complex mapping between priors on the neutrino masses to odds on the hierarchy, so a seemingly innocuous prior choice can strongly favor a particular hierarchy.

In this work we develop a methodology for a joint analysis of neutrino oscillation and cosmology data, which is agnostic to the hierarchy --- a hierarchy-agnostic prior. This guarantees that the relative odds of the NH:IH are driven by the data, and not by the choice of prior. We demonstrate using this prior that current data are not sufficiently constraining to determine the hierarchy. The limiting factor in determining the hierarchy is the neutrino mass sum measurement. We therefore set a target precision for future measurements of the neutrino mass sum, necessary in order to make a conclusive determination of the hierarchy. Assuming the NH minimum mass, this is sufficient for a conclusive determination, but other configurations will require an increased precision.

Current neutrino data is presented in Sec. \ref{sec:current_data}. The method for constructing a hierarchy-agnostic prior and calculating the relative odds NH:IH is described in Sec. \ref{sec:method}. The odds NH:IH for current data and a target precision for future experiments are presented in Sec. \ref{sec:results}, with conclusions summarized in Sec. \ref{sec:conclusions}. 

\section{Current Data} \label{sec:current_data}
Neutrino oscillation experiments have measured two mass-squared splittings: the mass-squared splitting between the closest two neutrinos $\Delta m_S^2$ (the small splitting) and the mass-squared splitting between the furthest and the midpoint of the closest two $\Delta m_L^2$ (the large splitting). If we label and order the masses as $m_a < m_b < m_c$, in the NH case we have
\begin{equation}
    \begin{aligned}
    &\Delta m_S^2 = m_b^2 -m_a^2 \ ,\\
    &\Delta m_L^2 = m_c^2 - \frac{1}{2}(m_b^2 +m_a^2) \ ,
    \end{aligned}
\end{equation}
and in the IH case,
\begin{equation}
    \begin{aligned}
    &\Delta m_S^2 = m_c^2 - m_b^2 \ , \\
    &\Delta m_L^2 = \frac{1}{2}(m_c^2 + m_b^2) -m_a^2 \ .
    \end{aligned}
\end{equation}
The current best constraints on these parameters are shown in Table \ref{tab:table1}. The constraint on the small splitting $\Delta m_S^2$ comes from combining data from the KamLAND \cite{KamLAND} experiment with a global analysis of solar, accelerator and short-baseline reactor neutrino experiments \cite{Gando:2013nba}. The constraint on the large splitting $\Delta m_L^2$ comes from a global analysis of data from atmospheric \cite{Aartsen:2017nmd, Abe:2017aap}, short-baseline reactor \cite{Adey:2018zwh, Bak:2018ydk} and long-baseline accelerator neutrino experiments \cite{Abe:2018wpn, NOvA:2018gge, Adamson:2014vgd}. Individual neutrinos are produced in interaction (i.e. flavor) eigenstates. Since the flavor eigenstates ($\nu_e$, $\nu_\mu$ and $\nu_\tau$) are mixtures of the mass eigenstates ($m_a$, $m_b$ and $m_c$), the neutrino flavor subsequently oscillates as it propagates. This means if a certain number of electron neutrinos ($\nu_e$) are produced by a source, as they propagate the number will change and manifest as a deficit of electron neutrinos and an excess of muon ($\nu_\mu$) and tau neutrinos ($\nu_\tau$). All of the experiments mentioned above search for a mismatch between the number of a particular neutrino flavor produced by a source and the number measured by the detector a certain distance away, using different sources and different distances. For example, KamLAND measures a deficit of electron anti-neutrinos using Japanese nuclear reactors as a source. 
\begin{table}[]
\caption{\label{tab:table1}%
Current best constraints for the mass-squared splittings with their associated 1$\sigma$ uncertainty from oscillation experiments  \cite{Splitting_constraints}, and 95\% CL upper bound on the sum of neutrino masses from cosmological data  \cite{Planck2018}.
}
\begin{ruledtabular}
\begin{tabular}{ll}
\textrm{Measurable Parameter} & \textrm{Current Constraint}\\
\colrule
\rule{0pt}{3ex}$\Delta m_S^2$ & \ $(7.53 \pm 0.18) \times 10^{-5} \ \mathrm{eV^2}$ \cite{Splitting_constraints}\\

$\Delta m_L^2$ \ (NH) & \ $(2.444 \pm 0.034) \times 10^{-3} \ \mathrm{eV^2}$ \cite{Splitting_constraints} \\

$\Delta m_L^2$ \ (IH) & \ $(2.53 \pm 0.05) \times 10^{-3} \ \mathrm{eV^2}$ \cite{Splitting_constraints}\\

$\summ$ & \ $< 0.12 \ \mathrm{eV}$ \cite{Planck2018}\\
\end{tabular}
\end{ruledtabular}
\end{table}

Cosmological data constrain the sum of neutrino masses:
\begin{equation}
    \summ =m_a + m_b + m_c \ ,
\end{equation}
and the current best constraint is shown in Table \ref{tab:table1} \cite{Planck2018}. This constraint comes from a combination of cosmological probes: the temperature and polarization fluctuations in the cosmic microwave background (CMB), which is the relic radiation from the surface of last scattering 380,000 years after the Big Bang \cite{Hu:2001bc}; weak gravitational lensing, which uses coherent distortions in observations of the CMB or galaxies to probe the matter distribution along the line of sight \cite{Bartelmann:1999yn, Lewis:2006fu}; and baryon acoustic oscillations (BAO), which measure a standard distance scale set by sound waves in the early universe \cite{Anderson:2013zyy}. The CMB and lensing part of this constraint comes from the \textit{Planck} satellite \cite{Planck2018}, and the BAO part comes from low redshift galaxy surveys \cite{Ade:2013zuv}. A larger neutrino mass sum suppresses the growth of structure in the universe; it is only by combining measurements of cosmic structure at early times, late times, small scales and large scales, that the effect of massive neutrinos can be determined.

\section{Method} \label{sec:method}
To quantify whether one hierarchy is favored over the other, we compute the posterior odds, given by
 \begin{equation} \label{eq:posterior_odds_expression}
    \frac{p(\mathrm{NH}|D)}{p(\mathrm{IH}|D)} = \frac{p(\mathrm{NH})}{p(\mathrm{IH})} \frac{p(D|\mathrm{NH})}{p(D|\mathrm{IH})} \ ,
\end{equation}
where \textit{D} represents current data. We wish to impose equal prior odds on the hierarchies, \textit{i.e.} $p(\mathrm{NH})= p(\mathrm{IH})$. Therefore, the first term on the right hand side is equal to one, and calculating the posterior odds equates to calculating the ratio of the marginal likelihoods, $p(D|\mathrm{NH})$ and $p(D|\mathrm{IH})$. Calculating the marginal likelihoods requires marginalizing over the individual neutrino properties, and we compute them via Monte Carlo integration:
 \begin{equation}\label{pDH}
    \begin{aligned}
    p&(D|\textrm{H}) = \int p(D | \boldsymbol{\theta}, \textrm{H}) p(\boldsymbol{\theta}|\textrm{H}) \ \textrm{d} \boldsymbol{\theta} \\
    &\approx \frac{1}{N} \sum_{i=1}^N p(\Delta m_S^2| \boldsymbol{\theta}_i, \mathrm{H}) p(\Delta m_L^2| \boldsymbol{\theta}_i, \mathrm{H}) p(\summ| \boldsymbol{\theta}_i, \mathrm{H}) \ .
    \end{aligned}
\end{equation}
Here $\boldsymbol{\theta}$ represents a parametrization which describes the properties of three neutrinos, $N$ is the number of sets of neutrino properties drawn from our prior $p(\boldsymbol{\theta}|\textrm{H})$, and $H$ is the hierarchy under consideration. The likelihood of the data given the parameter set and hierarchy, $p(D | \boldsymbol{\theta}, \textrm{H})$, can be split into the likelihoods $p(\Delta m_S^2| \boldsymbol{\theta}, \mathrm{H})$, $p(\Delta m_L^2| \boldsymbol{\theta}, \mathrm{H})$ and $p(\summ| \boldsymbol{\theta}, \mathrm{H})$ because the measurements of $\Delta m_S^2$, $\Delta m_L^2$ and $\summ$ are independent. In principle these individual likelihoods should be the data likelihoods from the experiments reported in Table \ref{tab:table1}. However, these results are independent and can be accurately approximated with surrogate likelihoods \footnote{Likelihoods which use the experimental measurements to approximate the data likelihoods.} for the purposes of this work. As such, the likelihoods for $\Delta m_S^2$ and $\Delta m_L^2$ are taken to be normal distributions with mean the measured value and standard deviation the associated uncertainty, given in Table \ref{tab:table1}. Note that $\Delta m_L^2$ differs between the hierarchies. Since cosmological data currently only places an upper bound on the sum of neutrino masses $\summ$, the likelihood is taken to be a normal distribution centered on zero with standard deviation half the 95\% upper bound.

If we draw $\boldsymbol{\theta}$ from a prior which favors the NH, the corresponding posterior odds will also be weighted in favor of the NH. For example, previous works have shown that defining $\boldsymbol{\theta}=\{m_a, m_b, m_c\}$ and drawing the three masses from the same log-normal distribution \footnote{The log-normal distribution is a continuous distribution where the logarithm of the variable is normally distributed.} is a seemingly reasonable choice which, however, strongly favors the NH \cite{Simpson:2017qvj, Gariazzo:2018pei}. We therefore require a prior which does not favor a hierarchy; where it is equally likely that a randomly drawn $\boldsymbol{\theta}$ corresponds to the NH as to the IH.

 Previous works have achieved this by adopting a continuous or discrete hierarchy parameter, which is positive for the NH and negative for the IH \cite{Jimenez:2010ev,Xu:2016ddc,Gerbino:2016ehw,Loureiro:2018pdz}. However, we achieve this through our choice of parametrization for $\boldsymbol{\theta}$. The only requirement is that the parameter set $\boldsymbol{\theta}$ can be translated to the measured quantities  $\Delta m_S^2$, $\Delta m_L^2$ and $\summ$. We choose $\boldsymbol{\theta} = \{\Delta m_a^2, \Delta m_b^2, m_a\}$ where $\Delta m_a^2$ is the mass-squared splitting between the lightest two neutrinos, $\Delta m_b^2$ the mass-squared splitting between the heaviest two, and $m_a$ the mass of the lightest neutrino. Explicitly
\begin{equation}
    \begin{aligned}
    &\Delta m_a^2 = m_b^2 -m_a^2 \ , \\
    &\Delta m_b^2 = m_c^2 - m_b^2 \ , \\
    \end{aligned}
\end{equation}
where $m_a < m_b < m_c$. Our approach means that the prior assumptions about neutrinos going into the analysis are explicit, which is not the case for \cite{Jimenez:2010ev, Xu:2016ddc}, and does not require introducing a separate discrete hierarchy parameter as in \cite{Gerbino:2016ehw,Loureiro:2018pdz}. Parametrizing neutrino properties in terms of a minimum mass and two splittings is quite common (e.g., \cite{Beacom:2002cb}, \cite{deSalas:2018bym, Gerbino:2016ehw}); however, the potential of this parametrization for constructing an explicitly equal-odds prior has not been previously investigated.

In the case of the NH $\Delta m_a^2 < \Delta m_b^2$ and the IH $\Delta m_b^2 < \Delta m_a^2$. Therefore, to construct a hierarchy-agnostic prior, we require $\Delta m_a^2 < \Delta m_b^2$ to be equally likely to $\Delta m_b^2 < \Delta m_a^2$. Hence we draw $\Delta m_a^2$ and $\Delta m_b^2$ from the same distribution. In our parametrization we wish the splittings to be positive and vary over a large range. There are a number of distributions which satisfy these requirements, for example uniform in log space (log-uniform) or log-normal. However, given the high accuracy of current splitting measurements, we have verified that the particular form of this distribution is unimportant. We therefore choose a log-normal distribution because it provides a proper prior \footnote{a prior distribution that integrates to 1}.

The parameter $m_a$ is required to translate $\Delta m_a^2$ and $\Delta m_b^2$ to the measurable quantities $\Delta m_S^2$, $\Delta m_L^2$ and $\summ$, as it provides an overall mass scale. Other possible options are $m_a^2$ or $\summ$. We require $m_a$ to be non-negative and little is known about its magnitude. In the case where a parameter can vary over many orders of magnitude a log-uniform prior is a reasonable choice. We found that our results were unchanged whether we used a log-uniform or a log-normal prior, so we once again choose a log-normal prior (see Appendix A).

The log-normal distributions used in this analysis are defined by the mean, $\mu$, and standard deviation, $\sigma$, of the underlying normal distribution. As such our prior space is defined by four parameters $\mu_s$, $\sigma_s$, $\mu_{m_a}$ and $\sigma_{m_a}$ for the priors on the splittings, $\Delta m_a^2$ and $\Delta m_b^2$, and $m_a$ respectively. Our final posterior odds result was found to be invariant over a large range of different choices for these prior parameters; in this work we specifically used $\mu_s = -9.25$, $\sigma_s = 5.0$, $\mu_{m_a} = 0.0$ and $\sigma_{m_a} = 7.0$, where $\Delta m_a^2$ and $\Delta m_b^2$ have units of $\mathrm{eV}^2$ and $m_a$ has units of eV. This choice gives a broad prior on the splittings because little is known about the magnitude of the mass-squared splittings before including oscillation data \cite{Splitting_constraints}. This choice is also motivated by the recent KATRIN result that the highest allowed value of the effective electron neutrino mass is less than 1.1 eV \cite{Aker:2019uuj}. Additionally, this specific choice translates into three approximately log-normal distributions on the individual neutrino masses, which are defined by distinct parameters. The structure of our hierarchy-agnostic prior means that these distributions are correlated (see Appendix B). This is in contrast to log-normal priors on the individual masses where the masses are drawn from the same distribution, which have previously been found to favor the NH \cite{Simpson:2017qvj, Gariazzo:2018pei}.

Once we have defined a hierarchy-agnostic prior, we randomly draw $N$ sets of neutrino properties, $\boldsymbol{\theta} = \{\Delta m_a^2, \Delta m_b^2, m_a\}$. Next, we use our samples of $\boldsymbol{\theta}$ to calculate $p(D|\mathrm{NH})$. The first step in computing $p(D|\mathrm{NH})$ is to translate the parameters $\Delta m_a^2$, $\Delta m_b^2$ and $m_a$ to the measurable quantities $\Delta m_S^2$, $\Delta m_L^2$ and $\summ$. Assuming the NH,
\begin{equation}
    \begin{aligned}
    &\Delta m_S^2 = \Delta m_a^2 \ , \\
    &\Delta m_L^2 = \Delta m_b^2 + \Delta m_a^2 / 2 \ ,\\
    \end{aligned}
\end{equation}
and 
\begin{equation}
\summ =m_a + \sqrt{m_a^2 + \Delta m_a^2} + \sqrt{m_a^2 + \Delta m_a^2 + \Delta m_b^2} \ .
\end{equation}
Once these values are found for every $\boldsymbol{\theta}$ they are used to compute the individual likelihoods $p(\Delta m_S^2| \boldsymbol{\theta}, \mathrm{NH})$, $p(\Delta m_L^2| \boldsymbol{\theta}, \mathrm{NH})$ and $p(\summ| \boldsymbol{\theta}, \mathrm{NH})$. Eq. (\ref{pDH}) is then employed to compute $p(D|\mathrm{NH})$.

Next we use the same set of $\boldsymbol{\theta}$ to compute $p(D|\mathrm{IH})$. This is allowable because our hierarchy-agnostic prior means that $p(\boldsymbol{\theta}|\textrm{NH}) =  p(\boldsymbol{\theta}|\textrm{IH})$. The procedure for computing $p(D|\mathrm{IH})$ is analogous to $p(D|\mathrm{NH})$. However, there are two differences: the translation of $\Delta m_a^2$ and $\Delta m_b^2$ to the measured splittings,
\begin{equation}
    \begin{aligned}
    &\Delta m_S^2 = \Delta m_b^2 \ , \\
    &\Delta m_L^2 = \Delta m_a^2 + \Delta m_b^2 / 2 \ ; \\
    \end{aligned}
\end{equation}
and the form of $p(\Delta m_L^2| \boldsymbol{\theta}, \mathrm{IH})$, as the measured value of $\Delta m_L^2$ differs between the hierarchies (see Table \ref{tab:table1}).

Once $p(D|\mathrm{NH})$ and $p(D|\mathrm{IH})$ are calculated, the ratio can be computed to find the posterior odds, as in Eq. (\ref{eq:posterior_odds_expression}). Since our calculation of the posterior odds is based on simulations, it is approximate. However, we find the posterior odds calculated with this method are approximately normal distributed if the number of sets of neutrino properties $N$ drawn in Eq. (\ref{pDH}) is equal to $10^9$. We can therefore use jackknife resampling to calculate the mean and variance. We use 100 sub-samples in our jackknife resampling, which gives us a mean accurate to the sub-percent level and a variance to the percent level when using $N=10^9$. Setting $N=10^9$ allows us to explore the sensitivity to various prior choices. 

\begin{figure*}
	\centering
	\includegraphics[width=0.7\textwidth]{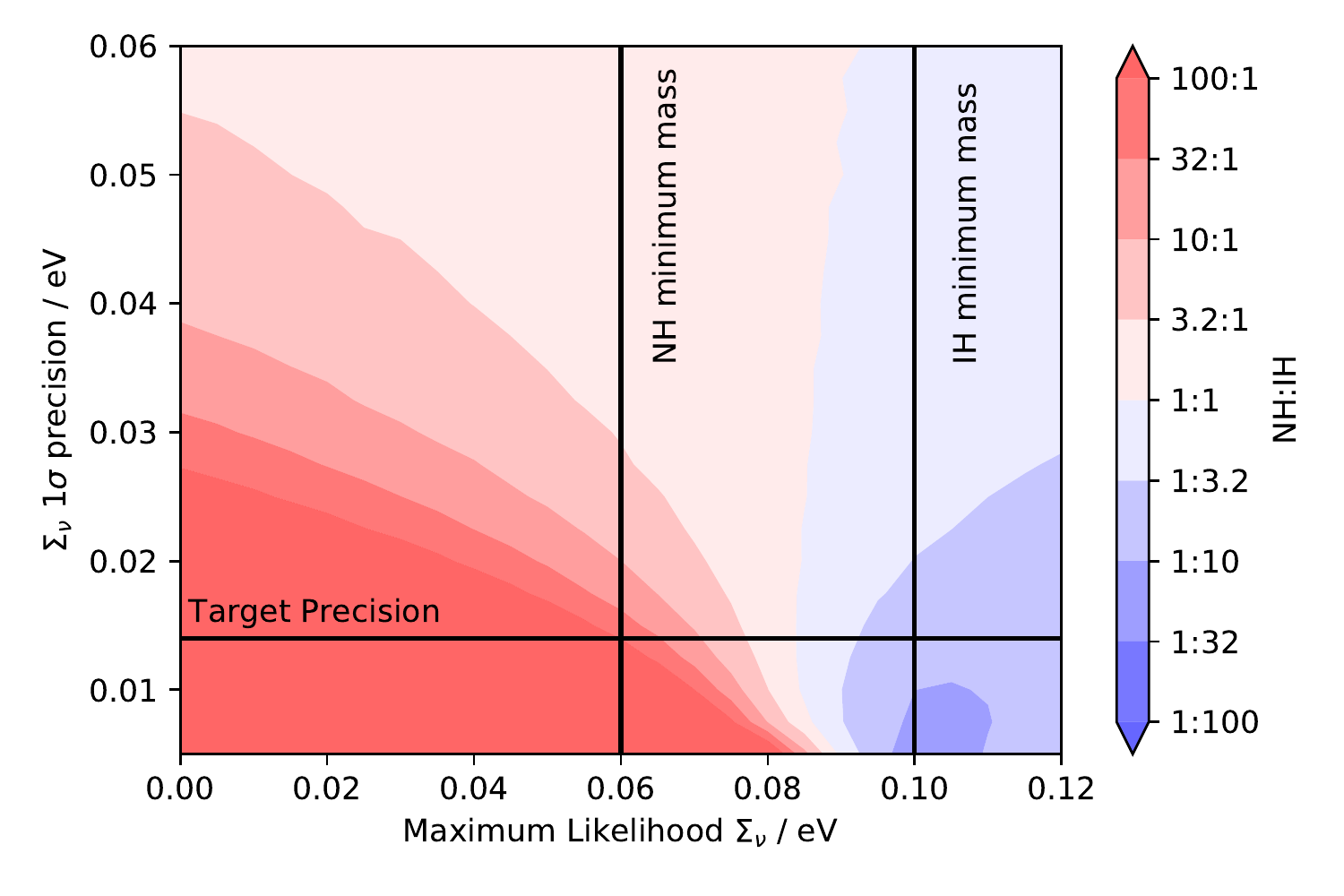}
	\caption{The dependence of the NH:IH posterior odds on the value of a future measurement of the neutrino mass sum ${\summ}$ and its associated 1$\sigma$ precision. The line at $\sigma({\summ})=0.014 \ \mathrm{eV}$ shows the target precision required to reach posterior odds of 100:1 in favor of the NH for a measurement at the NH minimum mass. The minimum masses are calculated from the current splitting measurements shown in Table \ref{tab:table1}.}
	\label{fig:measurement}
\end{figure*} 

\section{Results} \label{sec:results}
The posterior odds using our hierarchy-agnostic prior and current data are found to be
\begin{equation} \label{eq:posterior_odds}
    \frac{p(\mathrm{NH}|D)}{p(\mathrm{IH}|D)} = 2.66 \pm 0.04 \ ,
\end{equation}
where the 1$\sigma$ uncertainty is reported. Odds of 2.7:1 show a slight but inconclusive preference for the NH. This confirms that previous results strongly favoring the NH have been driven by the prior and not by the data \cite{Heavens:2018adv, Gariazzo:2018pei}.

We then turn to the question of what precision needs to be targeted by experiments in order to distinguish between the hierarchies. Since the splittings are much more accurately measured than the sum of the masses, we focus on setting a target for neutrino mass sum measurements. Figure \ref{fig:measurement} shows the dependence of the log posterior odds on the value of a future measurement of the neutrino mass sum ${\summ}$ and its associated 1$\sigma$ precision. The ${\summ}$ likelihood is assumed to be a normal distribution centred on the measured value. For a measurement at the NH minimum mass, 0.06 eV, the precision would need to be increased to $\sigma({\summ})$ = 0.014 eV to decisively determine the NH with odds of 100:1. A precision of 0.014 eV is an order of magnitude improvement on the current ${\summ}$ precision. This result does not change if we include future improved mass-squared splitting measurements from DUNE \cite{Dune} and JUNO \cite{Juno}, where the 1$\sigma$ precision on the mass-squared splittings is expected to improve by an order of magnitude over current results \cite{DUNE_forecasts, Juno_forecasts}.

However, we note that the preference for the IH, shown by the blue region on the right hand side of Fig. \ref{fig:measurement}, is driven by the prior on the lightest neutrino mass $m_a$ \cite{Gerbino:2016ehw,Agostini:2017jim,Caldwell:2017mqu}. This is because if the prior favors $m_a$ close to zero, but the sum of neutrino masses $\summ$ is measured to be close to the IH minimum mass 0.1 eV, the IH is favored over the NH. The preference for the IH (shown in Fig. \ref{fig:measurement}) disappears if one downweights smaller masses by using a uniform prior on $m_a$ instead of a uniform (or normal) prior on log $m_a$. As little is known about the order of magnitude of $m_a$, we chose to place a prior on log $m_a$. In spite of the prior sensitivity in this region, we have verified that our results -- the posterior odds for current data and the target precision for future experiments -- are robust to whether we use a prior on $m_a$ or log $m_a$. This demonstrates that our results are data-driven. See Appendix A for further details.

As mentioned previously, making a cosmological measurement of the neutrino mass sum requires combining multiple data sets at various epochs. Upcoming cosmological experiments which will contribute to a future measurement include: large scale structure (LSS) surveys such as the Large Synoptic Survey Telescope (LSST) \cite{LSST} and \textit{Euclid} \cite{Euclid}, which will use weak lensing and galaxy clustering to measure the matter distribution in the late-time universe; spectroscopic galaxy surveys such as the Dark Energy Spectroscopic Instrument (DESI) \cite{DESI} and the Maunakea Spectroscopic Explorer (MSE) \cite{MSE}, which will obtain a more accurate measurement of BAO and measure the matter distribution on smaller scales; ground-based CMB experiments such as the Simons Observatory \cite{Simons_ob} and CMB Stage-4 \cite{CMB_S4}, which will measure fluctuations in the temperature and polarization of the CMB on smaller scales; and the CMB satellite experiment LiteBIRD \cite{LiteBIRD}. A number of studies have predicted the ${\summ}$ precision which may be attainable with these experiments, and combinations thereof \cite{LSST_CMB, Euclid_def_study, Copeland_forecasts, DESI_forecasts, Font-Ribera:2013rwa, Babusiaux:2019spn_MSE, Simons_forecasts, DESI_CMB, Errard:2015cxa, Brinckmann:2018owf}. All of these analyses will be limited by systematic uncertainties in the ${\summ}$ measurement, not by statistical uncertainties. Therefore, our target precision of 0.014 eV imposes a stringent requirement on control of systematics in such analyses.

It is possible to measure the neutrino mass scale using data from particle physics as well as from cosmology. Current and upcoming experiments include: tritium beta decay experiments, such as KATRIN \cite{KATRIN} and \textit{Project-8} \cite{project-8}, which aim to measure the mass of the electron anti-neutrino; and neutrinoless double beta decay experiments, such as SuperNEMO \cite{supernemo} and KamLAND-Zen \cite{KamLAND-Zen:2016pfg}, which rely on neutrinos being their own anti-particle. These experiments involve completely different physics and systematics modeling from the cosmological constraints. They therefore may be vital in reaching our target precision \cite{Qian:2015waa, deSalas:2018bym}.

Current long baseline neutrino oscillation experiments, such as T2K \cite{T2K} and NOvA \cite{NOvA}, have some sensitivity to the mass hierarchy through electrons in matter affecting the neutrinos as they travel through the Earth \cite{Splitting_constraints}. These experiments have found a slight preference for the NH \cite{T2K_hierarchy, Nova_hierarchy}, which has not been included in this analysis. A possible extension of this work is to include this preference by scaling the surrogate likelihoods \cite{Heavens:2018adv}. One of the major goals of future oscillation experiments DUNE and JUNO is to determine the mass hierarchy \cite{DUNE_forecasts, Juno_forecasts}. These experiments are entering the construction phase but once built will have the sensitivity to provide an independent determination of the hierarchy, if sufficient neutrino interactions are recorded. 

\section{Conclusions} \label{sec:conclusions}
Combining data from neutrino oscillation experiments and cosmological probes has the potential to determine the neutrino hierarchy. In this work we developed a hierarchy-agnostic prior for a joint analysis of neutrino oscillation and cosmology data. Using current data we found odds for NH:IH of 2.7:1, which are inconclusive for determining the hierarchy. This is in contrast to previous analyses which found substantial odds in favor of the NH, demonstrating that these results were not data-driven.

We also computed the target precision, for future measurements of the neutrino mass sum, which is required to conclusively determine the NH with odds of 100:1. For a neutrino mass sum measurement at the NH minimum mass of 0.06 eV we found the precision on the neutrino mass sum would need to reach 0.014 eV. For higher masses the precision required is even greater. Future measurements of the neutrino mass sum will be systematics-limited. Therefore, our work demonstrates that an order of magnitude improvement in systematics control is needed to conclusively determine the neutrino hierarchy. While this is a challenging task, reaching this goal will be essential for our understanding of the Beyond Standard Model properties of the neutrino sector.

\begin{acknowledgments}
We thank John Beacom for a useful discussion about neutrino parametrizations, which simplified the construction of a hierarchy-agnostic prior. We also thank Andreu Font-Ribera, Will Handley, Alan Heavens and Andrew Pontzen for useful discussions.
CM was supported by the Spreadbury Fund, Perren Fund, and IMPACT Fund. 
BL was supported by NASA through the Einstein Postdoctoral Fellowship (award number PF6-170154).
HVP and JB were supported by the European Research Council (ERC) under the European Community's Seventh Framework Programme (FP7/2007-2013)/ERC grant agreement number 306478- CosmicDawn.
HVP was also supported by the research project grant ``Fundamental Physics from Cosmological Surveys" funded by the Swedish Research Council (VR) under Dnr 2017-04212.
AK was supported by STFC grant ST/S000666/1.
LC was supported by a Leverhulme Trust Research Project Grant.
RN was supported by STFC grant ST/N000285/1.
This work was partially supported by the UCL Cosmoparticle Initiative. 

\end{acknowledgments}

\appendix

\section{Prior on the lightest neutrino mass $m_a$}

Table \ref{tab:table2} shows how our results -- the posterior odds for current data and the target precision for future experiments -- are affected by the prior on the lightest neutrino mass ($m_a$). We find our results vary little when we place a uniform prior on $m_a$ instead of a log-normal prior as in the main text of this paper. Table \ref{tab:table2} also shows that our choice of a log-normal prior instead of a log-uniform prior on $m_a$ has little impact.

\begin{table}[H]
	\caption{\label{tab:table2}}
\begin{ruledtabular}
	\begin{tabular}{lll}
		\textrm{Prior on $m_a$ (eV)} & \textrm{Current Odds} & \textrm{Target Precision (eV)}\\
		\colrule
		\rule{0pt}{3ex}log-normal & $2.66 \pm 0.04$ & 0.014 \\
		uniform log, $10^{-10} - 1.1$ & $2.66 \pm 0.03$ & 0.014 \\
		uniform, $0.0-1.1$ & $2.83 \pm 0.14$ & 0.015 \\
	\end{tabular}
\end{ruledtabular}
\end{table}

\begin{figure*}
	\centering
	\includegraphics[width=0.7\textwidth]{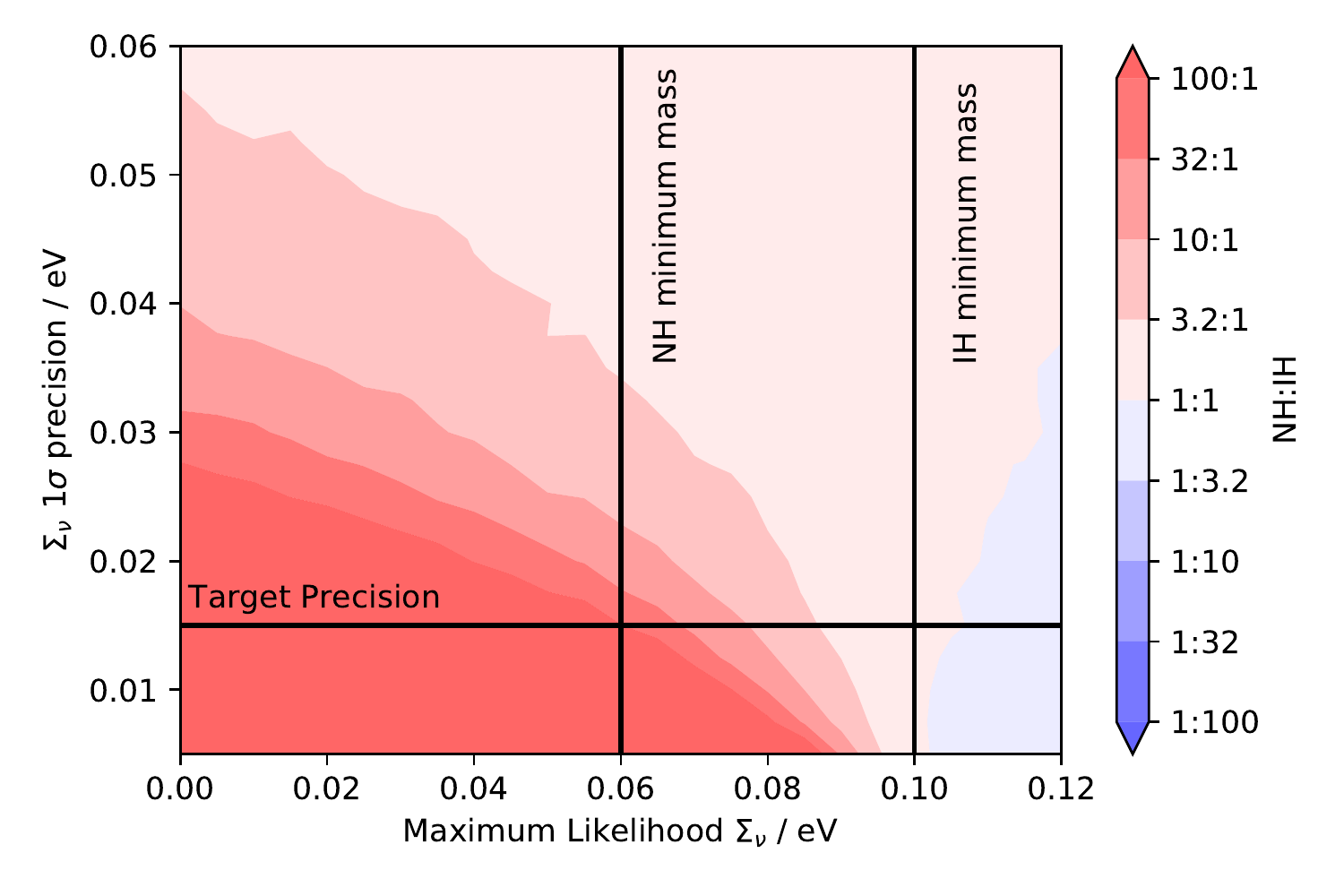}
	\caption{The dependence of the NH:IH posterior odds on the value of a future measurement of the neutrino mass sum ${\summ}$ and its associated 1$\sigma$ precision, with a uniform prior between $0.0 - 1.1$ eV on the lightest neutrino mass (instead of a log-normal prior as in Figure \ref{fig:measurement}).}
	\label{fig:uniform_prior}
\end{figure*}

Figure \ref{fig:uniform_prior} further illustrates that our results vary little when a uniform prior is placed on the lightest neutrino mass (the left hand side of the plot is consistent with Fig. \ref{fig:measurement}). However, it also shows that the preference for the IH shown on the right hand side of Fig. 1 disappears once a uniform prior is placed on the lightest neutrino mass. The preference for the IH is driven by the prior.

\section{Translation of Our Prior to Neutrino Mass Distributions}

\begin{figure*}
	\centering
	\includegraphics[width=0.8\textwidth]{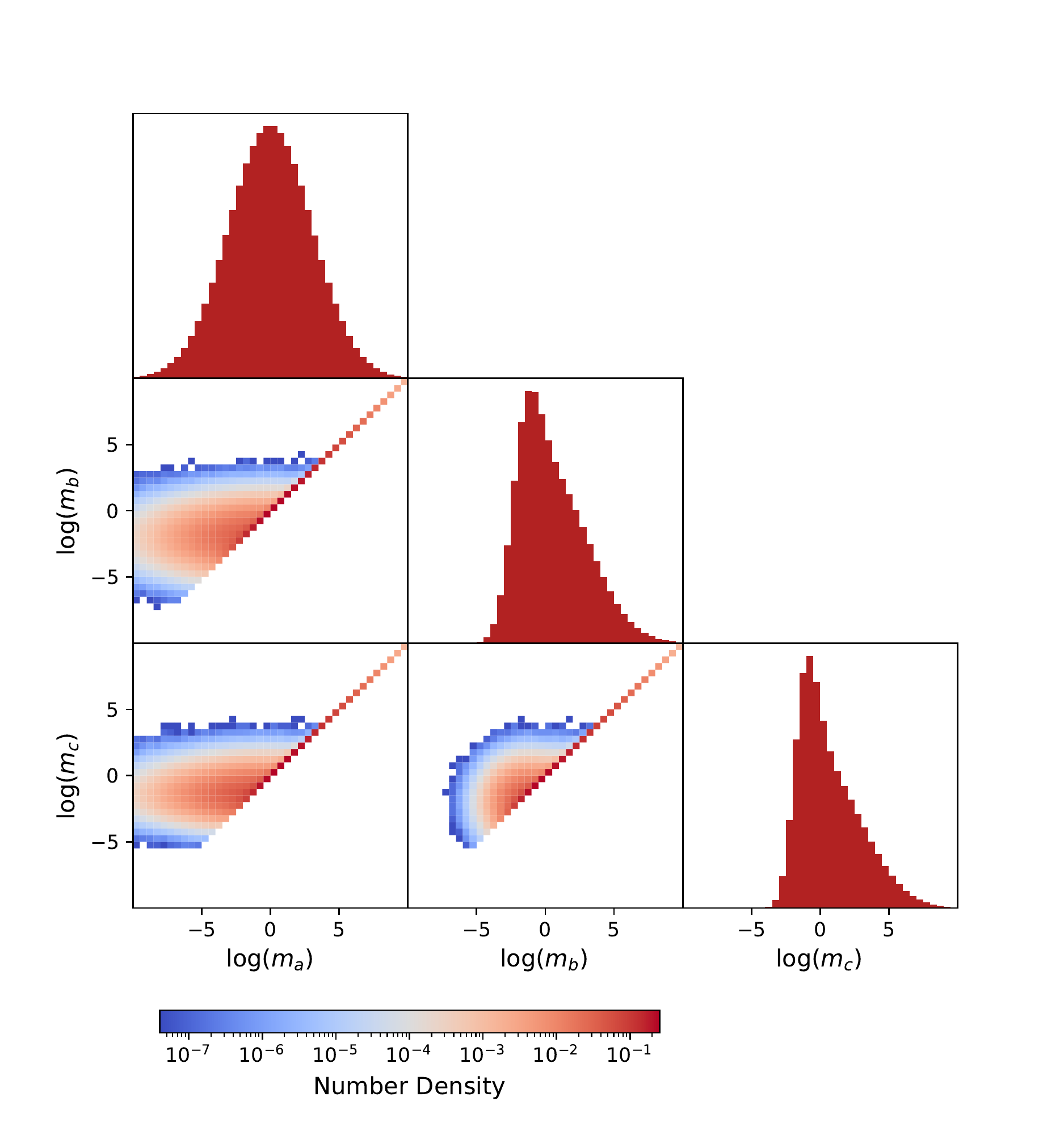}
	\caption{Our prior on the splittings, $\Delta m_a^2$ and $\Delta m_b^2$, and lightest neutrino mass, $m_a$,  translated to the log neutrino masses, where $m_a < m_b < m_c$.}
	\label{fig:masses}
\end{figure*}

The log normal priors on the splittings ($\Delta m_a^2$ and $\Delta m_b^2$) and lightest neutrino mass ($m_a$) used in this work translate into three approximately log-normal distributions on the individual neutrino masses ($m_a$, $m_b$, $m_c$), see Figure \ref{fig:masses}. These distributions are strongly correlated because $m_b$ and $m_c$ are computed by adding the splittings to $m_a$, so a larger $m_a$ results in a larger $m_b$ and $m_c$. As expected, they are particularly strongly correlated when $m_a$ is much larger than the splittings because the masses become quasi-degenerate. This pattern is also seen in neutrinoless double-$\beta$ decay discovery plots \cite{Agostini:2017jim, Splitting_constraints}.

\bibliographystyle{apsrev4-2}
\bibliography{apssamp} 

\end{document}